\documentclass[prc,preprint,aps,floatfix,fullpage]{revtex4}
\input{epsf}
\usepackage[dvips]{graphicx}
\def\dnuc{$^{12}_{\Delta}{\rm C} \,$}
\def\del{$\Delta \,$}

\begin{document}

\title{The $\Delta N \to NN$ transition in finite nuclei}
\author{C. Chumillas \footnote{e-mail: chumi@ecm.ub.es}, A. Parre\~no, A. Ramos}
\affiliation{Dept. ECM, Facultat de F\'{\i}sica, Univ. Barcelona,
E-08028, Barcelona, Spain
}
\date{\today}

\begin{abstract}
We perform a direct finite nucleus calculation of the partial width
of a bound $\Delta$ isobar decaying through the non-mesonic decay
mode, $\Delta N \to NN$. This transition is modeled by the
exchange of the long ranged $\pi$ meson and the shorter ranged
$\rho$ meson. The contribution of this decay channel is found to be
approximately 60 \% of the decay width of the \del particle in
free space. Considering the additional pionic decay mode, we
conclude that the total decay width of a bound $\Delta$ resonance 
in nuclei is of the order of 100 MeV and, consequently, 
no narrow \del nuclear states exist, contrary to recent claims 
in the literature. Our results
are in complete agreement with microscopic many-body calculations
and phenomenological approaches performed in nuclear matter.

\end{abstract}

\maketitle
\section{Introduction}

The $\Delta$(1232) isobar is a well established nucleon resonance
with spin-parity $J^\pi=\frac{3}{2}^+$ and isospin $I =
\frac{3}{2}$. This resonance has a width of 120 MeV in free space
from its strong decay to $\pi N$ states. Pion- and photo- nuclear
reactions at intermediate energies are dominated by the excitation
of the $\Delta$ in the nucleus. Within the nuclear environment,
the mesonic decay channel of a $\Delta$ gets reduced by the effect
of Pauli blocking, since the outgoing nucleon finds difficulties
in accessing nuclear unoccupied states. On the other hand, it is
well known that a $\Delta$ in the nucleus can also decay through
other mechanisms due precisely to the presence of the surrounding
nucleons. Among those, the most quantitatively important channel
being the one-nucleon induced process, $\Delta N \to NN$, which
leaves nucleons with enough momentum and energy to overcome the
Pauli blocking. This situation is analogue to what happens in
hypernuclear decay, where a $\Lambda$ particle bound in a nucleus
of mass 5 and beyond decays predominantly through the weak
$\Lambda N \to NN$ reaction. In the $\Delta N \to NN$ case, the
reaction is strong, and thus, with a much larger signal than in
the hypernuclear case.

In the late seventies and throughout the eighties, a lot of effort
was dedicated to understand the properties of the $\Delta$
resonance in the nuclear medium in order to describe pion-nucleus
scattering data
\cite{Wilkin:1973xd,Binon:1977ih,Clough:1974qt,Jansen:1978fj,
Ingram:1978cy,Albanese:1977zd,Piasetzky:1981ng,Altman:1983pb,Altman:1986nc}.
Many analyses were made in terms of the Delta-hole model, which
established an energy-dependent phenomenological $\Delta$
potential having an imaginary part of about $-40$ MeV at normal
nuclear matter density
\cite{Hirata:1977hg,Hirata:1977is,Hirata:1978wp,Horikawa:1980cv,Lenz:1982ac}. 
Microscopic
many-body calculations of the $\Delta$ width
\cite{Brown:1975di,Weise:1977ej,Oset:1979bi,Oset:1981ih,Oset:1987re,
Lee:1981st,Korfgen:1996ts,Hjorth-Jensen:1993rm},
including the reduction of the pionic decay mode, $\Delta \to \pi
N$, by the effect of Pauli blocking as well as its increase due to
the new non-mesonic mode, $\Delta N \to N N$, and other pion
absorption channels, obtain results that are in reasonable
agreement with the phenomenological potential and, hence, describe
satisfactory the pion-nucleus data. From all these studies it is
safe to assert that the in-medium $\Delta$ width stays within the
order of magnitude of the free width with a contribution from the
$\Delta N \to NN$ channel of about 40--70 MeV.

Having apparently reached a satisfactory description of the
$\Delta$ properties in a nuclear environment, 
the recent claims of the existence
of narrow $\Delta$ states in nuclei come as a surprise
\cite{Bartsch:1999ki}. The experiment was triggered by the
apparent existence of narrow $\Sigma$ states at CERN in spite of
the strong  $\Sigma N \to \Lambda N$ conversion mechanism
\cite{Bertini:1979qg,Bertini:1983nw}. A recent experiment
performed at Brookhaven with much better statistics did not
observe narrow states for targets of $^6$Li and $^9$Be in
$(K^-,\pi^{\pm})$ reactions, either for bound state or continuum
regions \cite{Bart:1999uh}, finalizing in this way many years of
debate and speculation over possible mechanisms that could explain
the existence of narrow $\Sigma$ states in nuclei
\cite{Oset:1989ey}. Nevertheless, some groups still claim the
question to be unsettled due to the limited energy resolution of
the recent experiments. The same groups advocate now the
possibility of finding narrow $\Delta$ states in nuclei, even if
the chances are {\it a priori} even worse than in the $\Sigma$
case due to the existence of the strong pionic decay mode that is
not completely blocked in a finite nucleus. These narrow $\Delta$
states in nuclei have also found some theoretical justification
\cite{Walcher:2001mv}, the rationale being that most of the former
theoretical works on the width of the $\Delta$ focused on a
kinematical situation appropriate for pion-nucleus scattering at
intermediate energies, where the $\Delta$ was created as a {\it
quasifree} state. This favored the overlap of its wave-function with 
those of the emitted nucleons, hence producing large values for the
decay width. However, the way the experiment of Ref.
\cite{Bartsch:1999ki} was devised, measuring pions and protons
in back-to-back coincidence, selects events in which the
$\Delta$ is {\it bound} in a nucleus which, according to the
theoretical model of Ref.~\cite{Walcher:2001mv}, could not decay
efficiently into two nucleons due to the little overlap between
the initial and final wave-functions.

The purpose of the present work is to perform a direct finite
nucleus calculation of the partial width of a {\it bound} $\Delta$
due to the non-mesonic mode $\Delta N \to NN$. Previous
calculations of this decaying mode in a finite nucleus either
concentrated on quasifree $\Delta$ states
\cite{Weise:1977ej,Oset:1979bi,Korfgen:1996ts} or were actually
induced from a finite-nucleus second order correction to nuclear
matter amplitudes \cite{Hjorth-Jensen:1993rm}.
 In order to account for all the physical
ranges of the strong $\Delta N$ interaction, we use a one-pion
plus one-rho exchange potential. While $\pi$-exchange is expected
to describe reasonably well the long and intermediate ranges
($r_\pi \sim (m_\pi)^{-1} \sim (140)^{-1}$ MeV$^{-1} \sim 1.4$
fm), $\rho$-exchange ($m_\rho \sim 770 $ MeV) is expected to cover
shorter distances. The rest of the members of the pseudoscalar and
vector octets (which would participate in a 
one-meson-exchange description of the process with masses up
to $\approx$ 1 GeV) are not included here, since they are forbidden
either by isospin conservation ($\eta, \omega$), or by flavor
conservation ($K, K^*$). For the description of the initial 
\dnuc \,nucleus we
 use a shell model, where the \del \, is assumed to weakly 
 couple to a 
$11$-particle core, from which we decouple the interacting
nucleon, leaving a (properly antisymmetrized) spectator system of
$10$ nucleons. To account for the effects of the strong
interaction in the initial two-body (\del $N$) state, we multiply
the corresponding (uncorrelated) two-body wave function, by an
appropriate correlation function that takes into account short-range repulsive
effects phenomenologically. As for the 
final two-nucleon state, we solve a $T-$matrix scattering
equation employing the Nijmegen nucleon-nucleon interaction.

The paper is organized as follows. In Sect.~\ref{formalism} we
present the formalism which allows us to write the nuclear
transition in terms of two-body matrix elements. In the same
Section we explicitly write the most general form for these
two-body amplitudes. In Sect.~\ref{OME} we build up the
regularized potential for the $\Delta N \to NN$ transition. In
Sect.~\ref{results} we present and discuss the results obtained,
and in Sect.~\ref{conclusions} we summarize our conclusions.

\section{Formalism}
\label{formalism}

\subsection{Decay rate}
\label{decayrate}

In the center-of-mass (CM) frame, the decay rate of a nucleus due to the non-mesonic
decay of a bound \del resonance, $\Delta N \to NN$, is given by:

\begin{equation}
\Gamma_{\rm nm} = \int  \frac{d^3 P}{(2\pi)^3}
 \int  \frac{d^3k}{(2\pi)^3}
\, \overline{\sum} \,\, (2\pi) \,\, \delta(M_I-E_R-E_1-E_2)
\mid {\cal M}_{fi} \mid^2 \ ,
\label{eq:rate1}
\end{equation}
where the initial bound system ($_\Delta^ A Z$ in what follows) has been considered
to be at rest. The quantity $M_I=M(A-1,Z) + M_{\Delta} - B_\Delta$ is the
mass of the initial nucleus with a
bound $\Delta$, with $B_\Delta$ being the $\Delta$ binding energy, while
$E_R$, $E_1$ and $E_2$ are
the energy of the residual \mbox{$(A-2)-$}particle system, and those of
the two emitted nucleons, respectively.
The integration variables ${\vec P}$ and ${\vec k}$ stand for
the total and relative CM momenta of the two nucleons in the final state.
The amplitude
${\cal M}_{fi} =
\left\langle \Psi_R ; {\vec P}{\vec k}\ s_1\,s_2\ t_1\,t_2
\right| \hat{O}_{\Delta N \to  NN}
\left| _\Delta^ A Z 
 \right\rangle$ 
corresponds to the transition from an  
initial nuclear state containing a $\Delta$ particle, to a final state which is
divided into two nucleons and a residual $(A-2)$-nucleon state, $\Psi_R$. 
The two-body
operator responsible for this transition has been designated by
$\hat{O}_{\Delta  N \to  NN}$.
The $\overline{\sum}$ sum indicates an average over 
the projections ($M_I$) of the initial nucleus total spin ($J_I$) 
and a sum over all quantum numbers ($J_R,M_R,T_R,T_{3_R}$)
of the residual $(A-2)-$particle system, as well as the spin ($s_1,s_2$)
and isospin ($t_1,t_2$) projections of the emitted nucleons. We follow Ref.
\cite{PaRaBe96} and adopt a weak-coupling scheme where a
$\Delta$ particle in an orbit $\alpha_\Delta = \{
n_\Delta,l_\Delta,s_\Delta,j_\Delta,m_\Delta \}$ and charge
$t_{3_\Delta}$ couples only to the ground-state wave function
of the nuclear $(A-1)$ core with quantum numbers $J_C,M_C,T_C,T_{3_C}$:
\begin{equation}
\mid \alpha_\Delta \rangle \otimes \mid A-1\rangle =
\sum_{m_\Delta \, M_C} \langle j_\Delta m_\Delta \, J_C M_C \mid
J_I M_I \rangle\, \mid (n_\Delta l_\Delta s_\Delta) j_\Delta
m_\Delta \rangle \, \mid J_C M_C \rangle \, \mid t_\Delta
t_{3_\Delta}\rangle \, \mid T_C T_{3_C} \rangle \ .
 \label{eq:coup}
\end{equation}

Employing the technique of the coefficients of fractional parentage, the core wave
function is further decomposed into a set of states where the nucleon in an orbit
$\alpha_N = \{ n_N,l_N,s_N,j_N,m_N \}$  is coupled to a residual
$(A-2)-$particle state:
\begin{eqnarray}
 \mid J_C M_C\, T_C T_{3_C} \rangle &=&
 \sum_{J_R T_R j_ N} \langle J_C\, T_C \{ \mid J_R \, T_R, j_N \,
 t_N
  \rangle \big[ \mid J_R,\, T_R \rangle \times \mid (n_ N l_N s_N)
j_N,\,  t_N \rangle \big]^{J_C M_C}_{T_C T_{3_C}} \nonumber \\
  &=&
 \sum_{J_R T_R j_N} \langle J_C\, T_C \{ \mid J_R \, T_R, j_N \,
 t_N  \rangle \nonumber \\
 &\times & \sum_{M_R\, m_N} \sum_{T_{3_R}\, t_{3_i}}
\langle J_R M_R\, j_N m_N \mid J_C M_C \rangle \langle
T_R T_{3_R}\, t_N t_{3_i} \mid T_C T_{3_C} \rangle
\nonumber \\
&\times& \mid J_R M_R \rangle \mid T_R T_{3_R} \rangle
\mid (n_N l_N s_N) j_N m_N \rangle \mid t_N t_{3_i} \rangle
\ ,
\label{eq:cfp}
\end{eqnarray}
where $t_N=\frac{1}{2}$. The appropriate spectroscopic factors 
$S = (A-1) \,\, \langle J_C T_C \{ \mid
J_R T_R , j_N t_N \rangle ^2$ in the case 
of $^{12}_\Delta$C 
are taken from Ref.~\cite{CoKu67}.

Assuming the $\Delta$ to decay from
 a $l_\Delta =0$ orbit,
 focusing only on the processes induced by the neutral $\Delta^0$
($t_\Delta=\frac{3}{2}$, $t_{3_\Delta}=-\frac{1}{2}$),
 and working
in a coupled two-body spin and isospin basis, the nonmesonic decay rate in
Eq. (\ref{eq:rate1}) can be written as:
\begin{equation}
\Gamma_{\rm nm}=\Gamma_{\rm n}+\Gamma_{\rm p} \ ,
\end{equation}
with $\Gamma_{\rm n}$ and $\Gamma_{\rm p}$ the neutron-
($\Delta {\rm n} \to {\rm nn}$) and proton-induced ($\Delta {\rm p}
\to {\rm np}$) decay rates, respectively, given by:
\begin{eqnarray}
\Gamma_{i} &=& \int \frac{d^3 P}{(2\pi)^3} \,
 \int \frac{d^3k}{(2\pi)^3} \,
(2\pi) \, \delta(M_I-E_R-E_1-E_2) \,
\sum_{S M_S} \sum_{J_R M_R} \sum_{T_R T_{3_R}}
\frac{1}{2J_I+1} \nonumber \\ &\times& \sum_{M_I}
\mid \langle T_R T_{3_R} \frac{1}{2} t_{3_i} \mid T_C T_{3_C}
\rangle \mid^2  \nonumber \\
&\times& \left| \,\,\sum_{T T_3} \langle T T_3 \mid \frac{1}{2}
t_1 \, \frac{1}{2} t_2 \rangle \sum_{m_\Delta M_C} \langle
j_\Delta m_\Delta \, J_C M_C \mid J_I M_I\rangle \sum_{j_N}
\, \sqrt{S \, ( J_C \, T_C \, ; J_R \, T_R \, , j_N \,
t_{3i} )}
\right. \nonumber \\
&\times&\sum_{M_R m_N} \langle J_R M_R\, j_N m_N \mid J_C M_C \rangle
 \sum_{m_{l_N} m_{s_N}} \langle j_N m_N \mid l_N m_{l_N} \frac{1}{2}
m_{s_N} \rangle  \nonumber \\
&\times& \sum_{S_0 M_{S_0}}
\langle S_0 M_{S_0} \mid \frac{3}{2} m_{\Delta} \, \frac{1}{2}
m_{s_N}
\rangle \,\, \sum_{T_0 T_{3_0}}
\langle T_0 T_{3_0} \mid \frac{3}{2}\,\, -\frac{1}{2} \,\,\,\,
\frac{1}{2} t_{3_i} \rangle \nonumber \\
&\times&  \left. t_{\Delta N \to  NN}
(S,M_S,T,T_3,S_0,M_{S_0},T_0,T_{3_0},l_\Delta,l_N,{\vec P},
{\vec k}) \phantom{\frac{1}{2}} \!\! \right|^2  \ , ~~~~i = {\rm n,p}
\label{eq:rate2}
\end{eqnarray}
where $t_{\Delta N \to NN}$ is the elementary $\Delta N \to NN$ transition
amplitude in the nucleus.


\subsection{Two-Body Amplitudes}
\label{2bamp}

In this section we derive the elementary two-body transition amplitude,
$t_{\Delta N \rightarrow NN}$, which describes the
one-nucleon induced decay of the $\Delta$-particle in nuclei.

First, we need to write the product of two single-particle wave
functions, \mbox{$\langle {\vec r}_1 \, \mid \alpha_\Delta
\rangle$} and $\langle {\vec r}_2 \, \mid \alpha_N \rangle$,
in terms of relative and center-of-mass coordinates,
${\vec r}$ and ${\vec R}$. Using the Moshinsky brackets~\cite{Mo59} one may
connect the wave-functions for two particles in a common harmonic
oscillator (H.O.) potential with the wave-function given in terms
of the relative and center-of-mass coordinates of the two
particles. In the present work, the single-particle $\Delta$ and $N$
orbits are taken to be solutions of harmonic oscillator mean field
potentials with parameters $b_\Delta$ and $b_N$
respectively. Their values are found by using the single-particle
energies of the \del particle given in Ref.~\cite{Bartsch:1999ki}
on the one hand, and by fitting the charge form factor of
$\phantom{,}^{12}$C on the other hand. The values
thus obtained are $b_\Delta=1.59$ fm and $b_N = 1.64$ fm
respectively. 
We assume the $\Delta$ binding energy to be given approximately by
the s-shell energy of the mean-field model used in Ref. \cite{Bartsch:1999ki},
namely $B_\Delta = -\varepsilon_{s_\frac{3}{2}} \sim 25$ MeV.

Assuming an average size parameter $b=(b_\Delta +
b_N)/2$ and working in the $LS$ representation, the product
of the two harmonic oscillator single-particle states,
$\Phi^{\Delta}_{nlm}({\vec r}_1 \, /b)$ and $\Phi^
N_{n'l'm'}({\vec r}_2 \, /b)$, can be transformed to a linear
combination of products of relative and center-of-mass wave
functions, $\Phi^{\rm rel}_{N_r L_r M_{L_r}}({\vec r} \, /\sqrt{2}
b)$ and $\Phi^{\rm CM}_{N_R L_R M_{L_R}}({\vec R} \, /
(b/\sqrt{2}))$, respectively. Since the $\Delta$ is in a
$l_\Delta=0$ shell, one obtains:
\begin{equation}
\Phi^{\Delta}_{100}\left(\frac{{\vec r}_1}{b}\right)
\Phi^N_{100}\left(\frac{{\vec r}_2}{b}\right) =
\Phi^{\rm rel}_{100}\left(\frac{{\vec r}}{\sqrt{2}b}\right)
\Phi^{\rm CM}_{100}\left(\frac{{\vec R}}{b/\sqrt{2}}\right)  \ ,
\label{eq:sshell}
\end{equation}
when the nucleon is in the s-shell and
\begin{eqnarray}
\lefteqn{
\Phi^{\Delta}_{100}\left(\frac{{\vec r}_1}{b}\right)
\Phi^N_{11m}\left(\frac{{\vec r}_2}{b}\right) = } \nonumber \\
& & \frac{1}{\sqrt{2}} \left\{ \Phi^{\rm
rel}_{100}\left(\frac{{\vec r}}{\sqrt{2}b}\right) \Phi^{\rm
CM}_{11m}\left(\frac{{\vec R}}{b/\sqrt{2}}\right)  - \Phi^{\rm
rel}_{11m}\left(\frac{{\vec r}}{\sqrt{2}b}\right) \Phi^{\rm
CM}_{100}\left(\frac{{\vec R}}{b/\sqrt{2}}\right) \right\}
\label{eq:pshell}
\end{eqnarray}
when the nucleon is in the p-shell.
With this decomposition, the amplitude
$t_{\Delta N \rightarrow NN}$ of \mbox{Eq. (\ref{eq:rate2})} can be
written in terms of amplitudes which depend on C.M. and relative
orbital angular momentum quantum numbers
\begin{eqnarray}
t_{\Delta  N \to  NN} &=
& \sum_{N_r L_r N_R L_R} X(N_r L_r N_R L_R, l_\Delta l_N)
\,\, t_{\Delta N \to NN}^{N_r L_r\, N_R L_R} \ ,
\label{eq:decamp}
\end{eqnarray}
where
$X(N_r L_r N_R L_R, l_\Delta l_N)$
are the Moshinsky brackets, which for $l_\Delta$=$l_N$=0 are just
$X(1\ 0\ 1\ 0,\ 0\ 0)=1$, and for $l_N=1$ are $X(1\ 0\ 1\ 1,\ 0\
1)=1/\sqrt{2}$ and
$X(1\ 1\ 1\ 0,\ 0\ 1)=-1/\sqrt{2}$.

As for the final $NN$ state, the antisymmetric state of
two independently moving nucleons with total momentum ${\vec P}$ and
relative momentum ${\vec k}$ reads:

\begin{equation}
\langle {\vec R}\, {\vec r} \, \mid {\vec P}\,{\vec k}\ S\,
M_S\ T\, M_T \rangle =
\frac{1}{\sqrt{2}} {\rm e}^{ {\rm i} {\vec P}\,{\vec R}} \left(
{\rm e}^{{\rm i} {\vec k}\,{\vec r}}
- (-1)^{S+T} {\rm e}^{- {\rm i} {\vec k}\,{\vec r}}
\right) \chi_{M_S}^S \chi_{M_T}^T  \ .
\label{eq:antwf}
\end{equation}

In order to incorporate the effects of the $NN$ interaction,
the plane wave describing the relative $NN$ motion needs to be
substituted by a distorted wave,
${\rm e}^{ {\rm i}{\vec k}\,{\vec r}} \to \Psi_{\vec k}({\vec r}\,)$,
solution of a $T-$matrix scattering equation, with the input of appropriate
and realistic baryon-baryon potentials.
The formalism to derive these distorted wave functions is described in
great detail in Ref.~\cite{AA01}. In the present work we strictly follow
such formalism while using the Nijmegen Soft-Core NSC97f strong potential model
\cite{StRi99} in
the calculation. The limited knowledge of the $\Delta N \to
\Delta N$ interaction, essentially due to the unknown $\Delta \Delta$-meson
vertices, has led us to treat the short-range $\Delta N$ effects
phenomenologically via the correlation function
\begin{equation}
f(r) = \left(1-{\rm e}^{-(r^2/a^2)}\right)^2 + b r^2 {\rm e}^{-(r^2/c^2)}
\ ,
\end{equation}
where $a=0.5$ fm, $b=0.25$ fm$^{-2}$ and $c=1.28$ fm. We have checked that our
results are rather insensitive to the particular shape and strength of this
correlation function, once realistic $NN$ wave-functions are incorporated.

The matrix elements
$t_{\Delta N \to  NN}^{N_r L_r\, N_R L_R}$ in Eq.~(\ref{eq:decamp}) are 
then given by:
\begin{eqnarray}
t_{\Delta N \to  NN}^{N_r L_r\, N_R L_R} &=
&
\frac{1}{\sqrt{2}}\int d^3R
\, \, \, \Phi^{\rm CM}_{N_R L_R}\left(\frac{{\vec R}}{b/\sqrt{2}}\right)
\, {\rm e}^{- {\rm i} {\vec P}\,{\vec R}} \nonumber \\
&\times& \int d^3r \, 
 \chi^{\dagger\, S}_{M_S}
\chi^{\dagger\, T}_{T_3} \, \Psi^*_{\vec k}\, ({\vec r}\,) \,
V_{\sigma \tau}({\vec r}\,) \,
f(r) \Phi^{\rm rel}_{N_r L_r}\left(\frac{{\vec r}}{\sqrt{2}b}\right)
\chi^{S_0}_{M_{S_0}} \chi^{T_0}_{T_{3_0}}\nonumber \\
&=& (2 \pi)^{3/2} \, 
\Phi^{CM}_{N_R L_R} \left({\vec
P}\frac{b}{\sqrt{2}}\right)
\,\, t_{\rm rel} \ ,
\label{eq:two}
\end{eqnarray}
with
\begin{equation}
t_{\rm rel}= \frac{1}{\sqrt{2}}
 \int d^3r \, \chi^{\dagger\, S}_{M_S}
\chi^{\dagger\, T}_{T_3} \,
\Psi^*_{\vec k} ({\vec r}\,) \,  V_{\sigma \tau}({\vec r}\,) \,
f(r) \Phi^{\rm rel}_{N_r L_r} \left(\frac{{\vec r}}{\sqrt{2}b}\right)
\chi^{S_0}_{M_{S_0}} \chi^{T_0}_{T_{3_0}} \ ,
\label{eq:trel}
\end{equation}
where, for simplicity, only the direct amplitude
--first term of Eq. (\ref{eq:antwf})-- is shown.
%

In the next section, we show how the potential
$V_{\sigma \tau}({\vec r}\,)$ can be decomposed as:
\begin{eqnarray}
V_{\sigma \tau}({\vec r}\,) & = & \sum_i \sum_\alpha V_\alpha^{(i)}(r)
 \hat{O}_\alpha\, \hat{I}  =  \sum_i \{
V_{SS}^{(i)}(r) \vec S_1 \vec \sigma_2 +
V_T^{(i)}(r) S_{12}(\hat r) \} \, \hat I
{\rm ,}
\end{eqnarray}
where the index $i$ runs over the different mesons exchanged ($\pi$ and $\rho$),
and $\alpha$ over the different spin operators, $\hat{O}_\alpha \in
( \
  {\vec S}_1 \,{\vec \sigma}_2 ,
\ S_{12}(\hat{r}) \equiv 3 \, {\vec S}_1 \, \hat{r} \, {\vec
\sigma}_2 \, \hat{r} - {\vec S}_1 \, {\vec \sigma}_2 )$,
written in terms of the spin $\frac{3}{2} \to \frac{1}{2}$ transition operator
$\vec S$ and the spin Pauli matrices $\vec \sigma$.
Since
both mesons have isospin 1, the isospin operator, $\hat{I}$,
factorizes out of the sum in $V_{\sigma \tau} ({\vec r}\,)$. This operator
takes the form $\vec T_1 \vec \tau_2$, with 
$\vec T$ and $\vec\tau$ having the
same structure as $\vec S$ and $\vec\sigma$, respectively. 
It can be shown that this operator 
only connects isotriplet $\Delta N$ and $NN$ states.

By performing a partial-wave expansion of
the final two-nucleon wave-function and working in the
$(LS)J$-coupling scheme,
the relative $\Delta N \to NN$
amplitude, $t_{\rm rel}$, can be further decomposed:
\begin{eqnarray}
t_{\rm rel} &=&
\frac{1}{\sqrt{2}} \,\,\sum_{i \, \alpha} \sum_{L L' J}
\,\, 4\pi {\rm i}^{-L'} \,\,
\langle L M_{L} S M_{S} | J M_J \rangle \,\, Y_{L M_L}
(\hat{k}_r) \nonumber \\
 & \times & \langle L_{r} M_{L_{r}} S_{0} M_{S_{0}} | J M_J \rangle \,\,
\langle (L' S) J M_J\mid \hat{O}_\alpha \mid (L_r S_0)  J M_J\rangle
\nonumber \\
 & \times & \langle T T_3 \mid \hat{I}\mid T_0 T_{3_0}\rangle
\, \int r^2 dr \, \Psi^{*\,J}_{L L'} (k_r,r) \, V_\alpha^{(i)}(r)
\, f(r)\,\Phi^{\rm rel}_{N_r L_r} (\frac{r}{\sqrt{2}b}) \ .
\label{eq:trel2}
\end{eqnarray}
The explicit expressions for
the expectation values of the spin dependent operator,
\mbox{
$\langle (L' S) J M_J\mid \hat{O}_\alpha \mid (L_r S_0)  J
M_J\rangle$},
can be found in the Appendix.

Taking into account all the possible initial states, the
required antisymmetry of the $NN$ wave function 
and the couplings induced by the $NN$ strong interactions,
the allowed
$\Delta N \to N N$ transitions 
are:

\begin{center}
\begin{tabular}{l c l}
$\phantom{.}^3 P_0 $ & $\to$ & $\phantom{.}^3 P_0 $ \\
$\phantom{.}^3 P_1 $ & $\to$ & $\phantom{.}^3 P_1 $ \\
$\phantom{.}^3 P_2 $ & $\to$ & $\phantom{.}^3 P_2 , \phantom{.}^3 F_2$ \\
$\phantom{.}^5 S_2 $ & $\to$ & $\phantom{.}^1 D_2 $ \\
\end{tabular}
\end{center}

\section{The Meson-Exchange Potential}
\label{OME}

\begin{figure}[ht]
\begin{center}
\includegraphics[width=4cm]{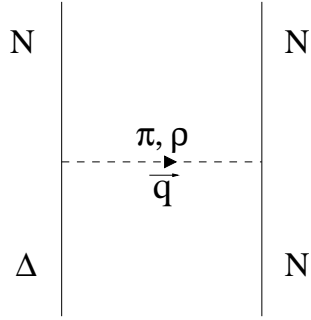}
\caption{Feynman diagram for the exchange of $\pi$ and $\rho$ mesons in the
$\Delta N \to NN$ transition.}
\label{fig:amp}
\end{center}
\end{figure}

As we already mentioned, we assume that the $\Delta N \to NN$ transition,
depicted in Fig.~\ref{fig:amp}, proceeds via
the exchange of the virtual $\pi$ and $\rho$ mesons.
The Lagrangians entering each vertex are~\cite{MaHoEl87}:
\begin{equation}
{\cal L}_{NN\pi}=\frac{f_{NN\pi}}{m_\pi} \bar \Psi
\gamma^5\gamma^\mu \vec \tau \Psi \partial_\mu \vec \Phi_\pi \,
{\rm ,} \label{LagNuNuPi}
\end{equation}
\begin{equation}
{\cal L}_{N\Delta\pi}=\frac{f_{N\Delta\pi}}{m_\pi} \bar \Psi \vec T \Psi_\mu \partial^\mu \vec \Phi_\pi \, {\rm ,}
\label{LagDelNuPi}
\end{equation}
\begin{equation}
{\cal L}_{NN\rho}=g_\rho \bar \Psi \gamma_\nu  {\bf \vec \tau}  \Psi  \vec \Phi^\nu_\rho
                  +\frac{f_\rho}{2 M} \bar \Psi \sigma_{\mu\nu}  \vec \tau \Psi \partial^\mu  \vec \Phi^\nu_\rho \, {\rm ,}
\label{lagNNR}
\end{equation}
and
\begin{equation}
{\cal L}_{N\Delta \rho}= {\rm i} \frac{f_{N\Delta\rho}}{m_\rho} \bar \Psi \gamma^5 \gamma _\nu  \vec T
\Psi_\mu (\partial^\nu  \vec \Phi^\mu_\rho-\partial^\mu  \vec \Phi^\nu_\rho) \, {\rm .}
\label{lagNDR}
\end{equation}
where $\Psi_\mu$ is the Rarita-Schwinger field operator describing the $\Delta$-isobar
and $M$ is the nucleon mass.

The nonrelativistic reduction of the free space Feynman amplitude
is associated with the transition potential. In momentum space the
one pion exchange potential takes the form [see Ref.~\cite{Brown:1975di}]:
\begin{equation}
V^{\pi}({\vec q}\,) = \frac{g_{NN\pi}}{2M} \frac{g_{N \Delta \pi}}
{2\overline{M}_{\Delta N}}\,
\frac{ ({\vec S}_1 \, {\vec q}\,) \,  ({\vec \sigma}_2 \, {\vec
q}\, )}{{\vec q}^{\; 2}+m_\pi^2-q_0^2} \,\, \vec T_1 \cdot \vec \tau_2 \ , \label{eq:pion}
\end{equation}
where ${\vec q} \,$ is the momentum carried by the pion directed towards the
$NN\pi$ vertex, $q_0$ its energy (which cannot be neglected due to the 
$m_{\Delta} - m_N$ difference)
 and $\overline M_{\Delta N}$
the average between the nucleon and $\Delta$ masses. 
We have introduced the coupling constants
$g_{ NN\pi}$ and $g_{N \Delta \pi}$ that relate to those in the
Lagrangians through ${f_{NN\pi}}/{m_\pi} = g_{NN\pi}/(2M)$ and
$f_{N\Delta\pi}/{m_\pi} = g_{N \Delta \pi}
/(2\overline{M}_{\Delta N})$, respectively.

For the $\rho$ meson
the potential takes the form:
\begin{equation}
V^{\rho}({\vec q}\,) = \frac{(g_\rho+f_\rho)}{2 M} \,\frac{g_{N \Delta
\rho}}{2 \overline M_{\Delta N}} \,
\frac{(\vec S_1 \times \vec q\,)\, (\vec \sigma_2 \times \vec q
\,)} {\vec q^{\; 2}+ m_\rho^2-q_0^2} \,\, \vec T_1 \cdot \vec \tau_2 \ , \label{eq:rhon}
\end{equation}
where $g_{N \Delta \rho}$ is defined through the relation 
${f_{ N\Delta\rho}}/{m_\rho}= g_{N \Delta
\rho}/({2 \overline M_{\Delta N}})$ .
Performing a Fourier transform of the general expression given in
Eq.~(\ref{eq:pion}) and Eq.~(\ref{eq:rhon}),
 using the relation $(\vec S_1 \times {\vec
q}\,)(\vec \sigma_2 \times {\vec q}\,)= (\vec S_1 \vec \sigma_2)
{\vec q}^{\, 2}- (\vec S_1 {\vec q}\,)(\vec \sigma_2 {\vec q}\,)
= \frac{2}{3} (\vec S_1 \vec \sigma_2) \vec q^{\,2}-\frac{1}{3} S_{12}(\hat q)\vec q ^{\,2}$,
it is easy to obtain
the corresponding transition potential in coordinate space, which
can be divided into central (SS) and tensor (T) pieces that take
the following form:
\begin{eqnarray}
V_{SS}^{(i)}(r) & = & K_{SS}^{(i)} \frac{1}{3}\left[
(m_i^2-q_0^2) \frac{e^{{\rm i}\sqrt{q_0^2-m_i^2}\, r}}{4 \pi r}
- \delta(r) \right] \equiv K_{SS}^{(i)} V_{SS}(r,m_i) \ ,
\\
V_T^{(i)}(r) & = & K_T^{(i)}\frac{1}{3}(m_i^2
-q_0^2)\frac{e^{{\rm i}\sqrt{q_0^2-m_i^2}\, r}}{4 \pi r} \left( 1 + \frac{3}{{\rm
i}\sqrt{q_0^2-m_i^2} r}-\frac{3}{(q_0^2-m_i^2) r^2} \right)
\equiv K_T^{(i)} V_T(r,m_i) \ , \nonumber
\end{eqnarray}
with $i = \mbox{$\pi$ or $\rho$}$.
In order to account for the finite size of the particles, we use a
monopole form factor $F_{i}({\vec q}^{\; 2})=
(\Lambda_i^2-m_i^2)/(\Lambda_i^2+{\vec q} ^{\; 2}-q_0^2)$
 at each vertex, where the value of the cut-off,
$\Lambda_i$, depends on the meson.
The use of form factors leads to the following
regularized potential for each meson:
\begin{eqnarray}
V_{\rm SS} (r; m_i) &\to& V_{\rm SS} (r; m_i) - V_{\rm SS}
(r;\Lambda_i) - \frac{1}{2} (\Lambda_i^2-m_i^2)
\sqrt{\Lambda_i^2-q_0^2} \, {\rm e}^{-\sqrt{\Lambda_i^2-q_0^2} \,
r} \left( 1-\frac{2}{\sqrt{\Lambda_i^2-q_0^2}r} \right) \, {\rm ,}
\nonumber \\
V_{\rm T} (r; m_i) &\to& V_{\rm T} (r; m_i) - V_{\rm T}
(r; \Lambda_i) -
\frac{1}{2} (\Lambda_i^2-m_i^2) \sqrt{\Lambda_i^2-q_0^2} \, {\rm
e}^{-\sqrt{\Lambda_i^2-q_0^2} \, r} \left(
1+\frac{1}{\sqrt{\Lambda_i^2-q_0^2}r} \right) \, {\rm .}
\nonumber \\
\label{eq:tregpot}
\end{eqnarray}
In Table~\ref{tab:constants} we show the explicit expressions for
the $K_\alpha^{(i)}$ coefficients, as well as the values of the
strong coupling constants and cutoffs used in this work.
\tabcolsep0.2cm
\begin{table}
\begin{center}
\caption{ 
Expressions for the $K_\alpha^{(i)}$ constants entering
the $\pi$ and $\rho$ potentials. The values of the
strong coupling constants and cutoffs are taken from Ref.~\cite{StRi99}.}
\vskip0.2cm
\label{tab:constants}
\begin{tabular}{ccccc}
\hline  \hline
\phantom{123} Meson \phantom{123}&
\phantom{1234} $K_{SS}^{(i)}$ \phantom{1234}
&
\phantom{1234} $K_T^{(i)}$ \phantom{1234}
&
\phantom{123} Strong CC \phantom{123}
&
\begin{tabular}{c}
$\Lambda_i$ \\
(GeV)
\end{tabular} \\
\hline
$\pi$ & $\displaystyle\frac{g_{ N\Delta\pi}}{2 \overline M_{\Delta N}}
\displaystyle\frac{g_{NN{\pi}}}{2 M}$ &
$ \displaystyle\frac{g_{N\Delta\pi}}{2 \overline M_{\Delta N}}
\displaystyle\frac{g_{NN\pi}}{2 M}$
&
$g_{NN\pi}=13.16$ & 1.3\\
 & & & $g_{N\Delta\pi}= 32.4$ & \\
\\
$\rho$
& $2\displaystyle\frac{g_{N\Delta\rho}}{2 \overline M_{\Delta N}}
\displaystyle\frac{f_\rho+g_\rho}{2 M}$ &
$-\displaystyle\frac{g_{N\Delta\rho}}{2 \overline M_{\Delta N}}
\displaystyle\frac{f_\rho+g_\rho}{2 M}$ &
$g_{N\Delta \rho}=38.1$ & 1.4\\
 & & & $f_\rho= 12.52$ & \\
 & & & $g_\rho=2.97$ & \\
\\ \hline \hline
\end{tabular}
\end{center}
\end{table}
Note that we use the phenomenological value
$\displaystyle\frac{f_{N\Delta\pi}}{f_{NN\pi}}=2.13$ 
which reproduces the free $\Delta$ width.
This same ratio is applied to obtain the coupling strength of the
$\Delta$ to the $\rho$ meson 
from that of the nucleon through the relation:
%
%
\begin{equation}
\frac{f_{N \Delta \rho}}{m_\rho} =
\left( \frac{f_{N \Delta \pi}}{f_{NN \pi}}\right) \,
\frac{g_{\rho}+f_{\rho}}{2 M}. 
\end{equation}

\section{Results}
\label{results}


Our results for the total decay rates are presented in
Table~\ref{tab:results} in units of the \del decay width in free
space, $\Gamma_\Delta = 120$ MeV, for the exchange of a $\pi$
meson, a $\rho$ meson and 
the combination of both.
The
different columns show results without strong short range correlations (label {\sl
free}), with only initial $\Delta N$ interactions
(label {\sl ISI}\,), and including, in addition, the
final $NN$ interactions (label {\sl ISI+FSI}\,) through the
corresponding strongly correlated wave functions
for the initial \del $N$ and final $NN$ states.
We also show separately the contribution coming from the decay
induced by a nucleon in the $l_N = 0$ shell (label {\sl
s}), by a nucleon in the $l_N = 1$ shell (label {\sl
p}) and the sum of both contributions (label {\sl
s+p}).

\tabcolsep0.2cm
\begin{table}[h!]
\begin{center}
\caption{Decay width of the \del isobar when the non-mesonic decay is modeled
by the exchange of a $\pi$ meson, a $\rho$ meson and
both mesons, $\pi + \rho$.
The results in the different columns correspond to not considering
strong short range correlations ({\sl free}), including only
the initial $\Delta N$
correlations ({\sl ISI}\,) 
or the combined effect of initial and final state interactions ({\sl ISI + FSI}\,).
Values are given in units of the free \del decay width,
$\Gamma_\Delta = 120$ MeV.}
\vspace*{0.3cm}
\label{tab:results}
 \begin{tabular}{|c|c|c|c|c|c|}
 \hline
shell & meson & {\sl free} & {\sl ISI} & {\sl ISI+FSI} \\
\hline
 & $\pi$ & 0.29 & 0.29 & 0.28 \\
 $s$ & $\rho$ & 0.22 $\times$ 10$^{-2}$ & 0.64 $\times$ 10$^{-4}$ &
 0.19 $\times$ 10$^{-3}$ \\
 & $\pi + \rho$ & 0.35 & 0.29 & 0.27  \\
 \hline
  & $\pi$ & 0.34 &  0.34 & 0.33 \\
 $p$ & $\rho$ & 0.48 $\times$ 10$^{-2}$ & 0.39 $\times$ 10$^{-3}$ &
 0.63 $\times$ 10$^{-3}$ \\
  & $\pi + \rho$ & 0.40 & 0.33 & 0.31 \\
  \hline
   & $\pi$ & 0.63 & 0.63 & 0.61 \\
 $s+p$ & $\rho$ & 0.69 $\times$ 10$^{-2}$ & 0.45 $\times$ 10$^{-3}$ &
 0.83 $\times$ 10$^{-3}$ \\
  & $\pi + \rho$  & 0.74 &  0.61 & 0.58 \\
 \hline
\end{tabular}
\end{center}
\end{table}

The results of Table~\ref{tab:results} show that the decay width is dominated
by the $\pi-$exchange mechanism. The $\rho-$exchange contribution is very small
although it interferes in a non-negligible way with $\pi-$exchange.
We also notice that short-range correlations
modify very  moderately the $\pi-$exchange decay width, representing 
a 3\% effect when both initial and final correlations are included.
Correlations affect in a much more relevant way
the $\rho-$exchange process, reducing the corresponding rate
by roughly a factor of 10.
In general, when introduced
independently, initial or final correlations reduce the total
$\pi+\rho$ decay rate by about 15\%, while their combined effect 
lowers the rate by 20\%.

\begin{figure}[ht]
\begin{center}
\includegraphics[width=9cm]{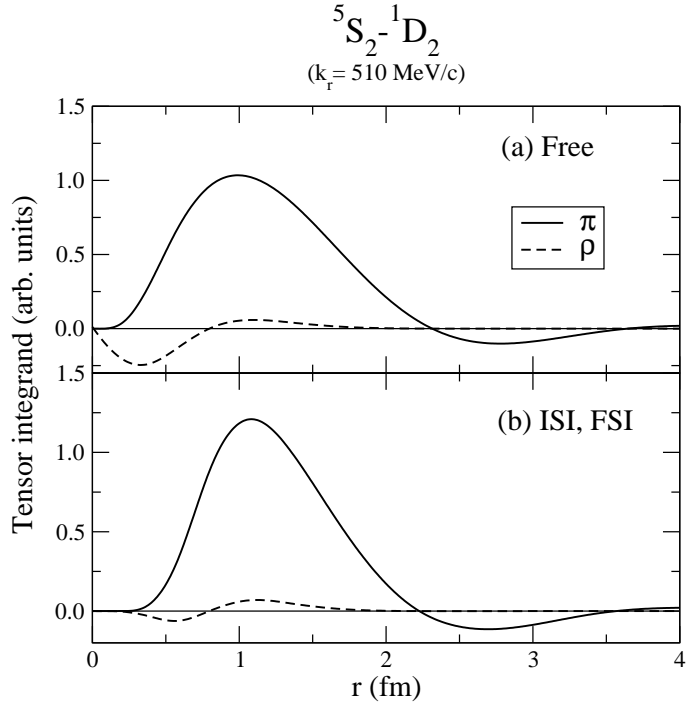}
\caption{$\pi$ (solid line) and $\rho$ (dashed line) contributions to
the integrand of the dominant $^5S_2 \to ^1D_2$ transition amplitude as a function
of the relative distance $r$. The upper panel displays the free amplitudes,
while the initial $\Delta N$ ({\sl ISI}) and final $NN$ ({\sl FSI}) short-range effects
are included in the amplitudes of the lower panel.}
\label{fig:pirho}
\end{center}
\end{figure}

The systematics observed in the results presented in Table~\ref{tab:results}
can be better understood by examining
Fig.~\ref{fig:pirho}. There, the integrand of the tensor
transition amplitude in Eq.~(\ref{eq:trel2}), 
which gives the most important contribution to the
$\Delta$ width as we will see,  
is shown as a function of the relative distance $r$,
for a representative relative $NN$ momentum value of $|\vec k_r|= 510$ MeV/c
and in the case
of the $\pi-$exchange (solid lines) and the
$\rho$-exchange (dashed lines) mechanisms. All the results include
form factors at the
vertices, but the amplitudes in the upper panel are obtained
using uncorrelated $\Delta N$ and $NN$ wave functions, while
both initial and final correlations are included in the results of
the lower panel.
One clearly sees that, being of long range nature, the pion-exchange amplitude
is little affected by the inclusion of short-range correlations. In contrast,
the $\rho-$exchange potential, being more short ranged due to
its larger meson mass, peaks at much shorter distances, hence the 
implementation of strongly correlated wave functions in the lower panel  reduces the
integrand substantially. In addition, the $\rho-$exchange
potential shows a node in the space region of relevance, which induces an
additional
cancellation of the $\rho-$contribution once {\sl ISI+FSI}  effects are included.

The contribution of the tensor transition potential to
the decay width dominates by two orders of magnitude that
of the central term, as we can see in Table~\ref{tab:results2}.
 This is essentially due to the
fermionic character of the final two-nucleon state which requires
the $NN$ wave function to be antisymmetric, {\it i.e.}, the relative
$NN$ quantum numbers have to verify the $L+S+T=$odd relation, where
$L, S$ and $T$ stand for angular momentum, spin, and isospin
respectively. Since $T$ is necessarily 1 and the central spin
operator connects only states with $S_0=S=1$, the orbital angular
momentum $L$ of the final $NN$ must be odd. This can only be
achieved from an initial $\Delta N$ pair having $L_r=1$, which
means that the central transition amplitude induced by s-shell
nucleons is exactly zero since we consider the $\Delta$ to be in
the lowest energy state, $1s_\frac{1}{2}$. For p-shell nucleons the
central transition is not forbidden, since 50\% of the wave function has
a relative angular momentum $L_r=1$, but even in this case the tensor
transitions dominate due to the large momentum transferred in the reaction
($|\vec {\rm q}| \sim 500$ MeV/c), which favors also a large
amount of angular momentum transfer.

\tabcolsep0.2cm
\begin{table}[h!]
\begin{center}
\caption{Decay width of the \del isobar when the non-mesonic decay is modeled
by the exchange of a $\pi+\rho$ mesons 
considering
initial and final state interactions ({\sl ISI + FSI}\,).
The different contributions of the central and tensor channels are
presented and their contributions to the diferents shells. Values are given in units of the free \del decay width,
$\Gamma_\Delta = 120$ MeV.}
\vspace*{0.3cm}
\label{tab:results2}
 \begin{tabular}{|c|c|c|c|}
 \hline
shell & Central & Tensor & TOTAL \\
\hline
$s$   & 0 & 0.27 & 0.27\\
$p$   & 0.16 $\times 10^{-2}$& 0.31& 0.31\\
$s+p$ & 0.16 $\times 10^{-2}$ & 0.58 & 0.58\\
 \hline
\end{tabular}
\end{center}
\end{table}

The calculated in-medium width of the $\Delta$ due to the
non-mesonic $\Delta N \to NN$ mechanism represents a 58\% fraction
of the free width, i.e. it amounts to $\Gamma_\Delta=70$ MeV. In
the optical potential language this would correspond to an imaginary part
of about ${\rm Im}\, U_{\Delta}= \Gamma_\Delta/2 = 35$~MeV, in
perfect agreement with extrapolations at zero momentum of the
absorptive optical potential, calculated for quasifree $\Delta$'s
in Ref. \cite{Lee:1981st}, and with the phenomenological analysis in various
nuclei \cite{Hirata:1977hg,Hirata:1977is,Hirata:1978wp,Horikawa:1980cv,Lenz:1982ac}. 

Our result is about 50\% larger than
the finite nucleus calculation of Ref.~\cite{Hjorth-Jensen:1993rm}, where
the effect of $NN$ correlations is accounted for via a nuclear-matter G-matrix
while direct $\Delta N$ short range correlations are ignored.
The origin of the discrepancy comes essentially from the use of a
phenomenological $N\Delta \pi$ coupling constant here, with a value
$f_{\pi N\Delta}/f_{\pi NN} = 2.13$ adjusted to reproduce the free $\Delta$ width,
which is 25\% larger than the the quark model value 
$f_{\pi N\Delta}/f_{\pi NN} = 6\sqrt{2}/5=1.70$ employed in 
Ref.~\cite{Hjorth-Jensen:1993rm}. 

Finally, we
also note that the contribution to the
$\Delta$ width from the $\Delta N\to NN$ mechanism explored here is
in excellent agreement with the analogous
nucleon pole contribution calculated by Oset and Salcedo \cite{Oset:1987re}
in nuclear matter for a pion kinetic energy of $T_{\pi} \sim 100$ MeV,
which, in their notation, would correspond to a
$\Delta$ binding energy of 25 MeV, as assumed in the present work,
although with a finite momentum of 150--200 MeV/c.

\section{Conclusions}
\label{conclusions}

Motivated by recent speculations on the possible existence of narrow
$\Delta$-nuclear states, we have performed the first direct finite nucleus
calculation of the partial width of
a {\it bound} $\Delta$ resonance via the decay mechanism $\Delta N \to NN$,
including $\Delta N$ correlations and realistic $NN$ interactions.
We find that, in $^{12}_\Delta$C, this partial width represents a 58\% fraction
of the free width, i.e. it amounts to $\Gamma_\Delta=70$ MeV.

Our result, evaluated explicitly for a $\Delta$ nuclear bound state, is 
in quantitative agreement
with extrapolations to low momentum of the  partial widths obtained
for $\Delta$  quasifree states in nuclear matter.

Considering also the partial width of the mesonic mode $\Delta
\to N \pi$, which can be quenched by about 50\% in nuclei due to Pauli
blocking, we therefore conclude that the total decay width of a {\it bound} $\Delta$
resonance in nuclei is of the order of 100 MeV and, consequently, narrow
$\Delta$ states cannot be formed in finite nuclei.

\section{Acknowledgments}
\label{acknowledgments}

We are grateful to A. Polls for very enlightening discussions.
This work is partly supported by
contract FIS2005-03142 from MEC (Spain) and FEDER
and by the Generalitat de Catalunya under contract 2005SGR-00343.
This research is part of the EU Integrated Infrastructure Initiative
Hadron Physics Project under contract number RII3-CT-2004-506078.

\setcounter{equation}{0}
\setcounter{section}{0}

\renewcommand{\theequation}{\Alph{section}.\arabic{equation}}
\setcounter{section}{1}
\setcounter{equation}{1}
\section*{APPENDIX}

In this Appendix, the explicit expressions for the
$\langle (L' S) J M_J | {\hat O}_\alpha | (L_r S_0) J M_J \rangle$
coefficients
appearing in the evaluation of the relative $\Delta N \rightarrow NN$
amplitude will
be given. The quantum numbers $L_r, S_0, J$ and $M_J$ stand for the
initial relative orbital angular momentum, the initial coupled intrinsic spin and
the total spin and spin projection of the $\Delta N$ state, while the numbers $L',S,J$ and
$M_J$ are the pertinent quantities for the final $NN$ system.

\subsection{Spin-Spin Transition}

\begin{eqnarray}
& & {\hat O}_\alpha = {\vec S_1} {\vec \sigma_2}
\nonumber \\
& & \langle (L' S) J M_J | {\hat O}_\alpha | (L_r S_0) J M_J \rangle
= -\frac{4}{\sqrt{6}}\,\,\, \delta_{L_r L'} \delta_{S_0 S} \delta_{S 1}
\label{coefss}
\end{eqnarray}

\subsection{Tensor transition}

\begin{eqnarray}
& & {\hat O}_\alpha = S_{12} ({\hat r}) = 3 {\vec S}_1 {\hat r}
{\vec \sigma}_2 {\hat r} -  {\vec S_1}{\vec \sigma}_2
\label{coeftens}
\end{eqnarray}
The tensor operator only allows for $S_0=2 \to S=0$ and $S_0=1 \to
S=1$ transitions, with matrix elements:

\vskip0.5cm

\begin{center}
\tabcolsep0.1cm
\begin{tabular}{|c|c|c|c|c|c|}
\hline
 & & & & & \\
 $ S_0=2 \to S=0$ & $L_r=J+2$ & $L_r=J+1$ & $L_r=J$ & $L_r=J-1$ & $L_r=J-2$ \\
 & & & & & \\ \hline
  & & & & & \\
$L^\prime = J$ & $-3 \sqrt{\frac{(J+1)(J+2)}{(2J+1)(2J+3)}} $ &
0&
$\sqrt{\frac{6 J (J+1)}{(2J-1)(2J+3)}}$&
0 &
$-3\sqrt{\frac{ J (J-1)}{(2J+1)(2J-1)}}$
\\
 & & & & & \\
\hline
\end{tabular}
\end{center}

\vskip1cm
\begin{center}
\begin{tabular}{|c|c|c|c|}
\hline
 & & & \\
 $S_0=1 \to S=1$  & $L_r=J+1$ & $L_r=J$ & $L_r=J-1$ \\
 & & & \\ \hline
  & & & \\
$L^\prime = J+1$ & $-\frac{J+2}{\sqrt{6}(2J+1)}$ &
0&
$
\frac{3}{2J+1}\sqrt{\frac{J(J+1)}{6}}
$
\\
 & & & \\ \hline
& & & \\
$L^\prime = J $ & 0 &
$\frac{1}{\sqrt{6}}$&
0
\\
 & & & \\ \hline
& & & \\
$L^\prime = J -1$ &
$\frac{1}{2J+1}\sqrt{\frac{3}{2}J (J+1)}$&
0
&
$-\frac{1}{\sqrt{6}} \left( \frac{J-1}{2J+1} \right)$
\\
 & & & \\
\hline
\end{tabular}
\end{center}


\subsection{Isospin matrix elements}

\begin{eqnarray}
& & {\hat I} = {\vec T}_1 {\vec \tau}_2
\nonumber \\
& & \langle T T_3 | {\hat I} | T_0 T_{30} \rangle
= -\frac{4}{\sqrt{6}}\,\,\, \delta_{T_3 T_{30}} \delta_{T_0 T} \delta_{T 1}
\label{coefsT}
\end{eqnarray}

\end{document}